\documentclass[pr,twocolumn,showpacs]{revtex4}
\usepackage{amssymb}
\usepackage{amsmath}
\usepackage{graphicx}
\usepackage{bm}
\usepackage{color}
\usepackage{ulem}
\usepackage[T2A]{fontenc}

\newcommand{\eps}{\varepsilon}

\renewcommand{\i}{{\mathrm i}}
\newcommand{\braket}[3]{\left\langle #1 \left| #2 \right| #3 \right\rangle}
\newcommand{\ket}[1]{\left| #1 \right\rangle}
\newcommand{\dz}[0]{\frac{d}{dz}}

\begin{document}

\title{Zeeman splitting of light hole in quantum wells: comparison of theory and experiments}
\author{M. V. Durnev}
\affiliation{Ioffe Physical-Technical Institute of the RAS, 194021 St. Petersburg, Russia}

\begin{abstract}
The theory for light-hole Zeeman splitting developed in [Physica E 44, 797 (2012)] is compared with experimental data found in literature for GaAs/AlGaAs, InGaAs/InP and CdTe/CdMgTe quantum wells. It is shown that the description of experiments is possible with account for excitonic effects and peculiarities of the hole energy spectrum in a quantum well including complex structure of the valence band and the interface mixing of light and heavy holes. It is demonstrated that the absolute values and the sign of the light-hole $g$-factor are extremely sensitive to the parametrization of the Luttinger Hamiltonian.

\end{abstract}

\maketitle
\section{Introduction}

Size quantization of charge carriers in semiconductor nanostructures leads to a considerable renormalization of the $g$-factor of electrons and holes~\cite{ivchenko_kiselev92, PhysRevB.75.245302, kiselevmoiseev96_rus, PhysRevB.50.8889, Durnev2012797}. This renormalization is particularly noticeable for the light hole in quantum wells due to magneto-induced mixing of the light-hole ground state ($lh1$) and the first excited state of a heavy hole ($hh2$) which was predicted in the theoretical work~\cite{Durnev2012797}. Large values of the light-hole $g$-factor ($g_{lh1}$) were also observed in experiments including measurements of the magneto-photoluminescence spectra~\cite{timofeev, PhysRevLett.79.3974,petrov_ivanov}, quantum beatings between the excitonic spin sub levels~\cite{PhysRevB.48.1955}, magnetotransmission~\cite{PhysRevB.55.9924} and magnetoabsorption~\cite{10.1063/1.2245213} spectra, and magneto-optical Kerr effect~\cite{arora:213505}. The values of the light-hole $g$-factor obtained in the mentioned experiments (see Tab.~\ref{table1}) considerably exceed those for an electron and a heavy hole observed in GaAs/AlGaAs-type wells. It is noteworthy that an experimental determination of the $g_{lh1}$ sign appears to be a hard task therefore we will further analyze only its absolute values.

This work is aimed to a detailed analysis of the experimental data on the light-hole $g$-factor in GaAs/AlGaAs, InGaAs/InP and CdTe/CdMgTe quantum wells and its comparison with theory. Theoretical model of Ref.~\cite{Durnev2012797} based on the resonant two-level approximation (one considers $lh1$ and $hh2$ subbands only) is extended including magneto-induced mixing of the state under study  with  all hole states of both discrete and continuum spectra in the framework of the Luttinger Hamiltonian. We have also developed a theory for Zeeman splitting of the hole in the vicinity of the critical width (well width at which the crossing of $lh1$ and $hh2$ levels occurs) with account for the interface mixing of heavy and light holes which predicts a nonlinear in magnetic field contribution to Zeeman effect. The theory of Zeeman effect for a light-hole exciton including the interface mixing is presented. It is shown that the Coulomb interaction between an electron and hole results in a linear in magnetic field splitting of the exctionic spin sub levels even in wells of the critical width. The results of calculations including excitonic effects are in a reasonable agreement with the experimental data.

\begin{table}
\caption{Experimental data on the light-hole $g$-factor to be analyzed in this work}
 \begin{tabular}{p{0.5\linewidth}p{0.3\linewidth}c}
\hline
\hline
Material & Well width & $g_{lh1}$ \\
\hline
GaAs/Al$_{0.3}$Ga$_{0.7}$As & $30$~\AA~\cite{PhysRevB.48.1955} & 1.4$^*$ \\
 & $120$~\AA~\cite{PhysRevB.48.1955} & 2.9$^*$ \\
  & $150$~\AA~\cite{timofeev} & 4$^*$\\
\hline
GaAs/Al$_{0.36}$Ga$_{0.64}$As & $20\div180$~\AA~\cite{10.1063/1.2245213} & 2$\div$6 \\
 & $180$~\AA~\cite{petrov_ivanov} & -9.4 \\
\hline
GaAs/Al$_{0.33}$Ga$_{0.67}$As & $43\div140$~\AA~\cite{arora:213505} & 6$\div$9 \\
\hline
In$_{0.53}$Ga$_{0.47}$As/InP & $100$~\AA~\cite{PhysRevB.55.9924} & 8.9$\pm$1.2 \\
\hline
CdTe/Cd$_{0.74}$Mg$_{0.26}$Te & $75$~\AA~\cite{PhysRevLett.79.3974, PhysRevB.56.2114} & -3 \\
\hline
\hline
\end{tabular}
\vskip0.2cm
\begin{flushleft}
$^*$ Absolute value of $g_{lh1}$
\end{flushleft}
\label{table1}
\end{table}
  
\section{Model}

Longitudinal Zeeman effect for holes (magnetic field applied along the quantum well growth axis ${\bm B} \parallel z \parallel [001]$) consists of two contributions. The first one corresponds to a bulk hole and can be described with the following Hamiltonian
\begin{equation}
\label{eq:bulk}
{\cal H}_B = -2\varkappa \mu_B {\bm J}\cdot{\bm B}\:,
\end{equation}
where $\varkappa$ is the magnetic Luttinger parameter, $\mu_B$ is the Bohr magneton, and ${\bm J}$ is the pseudo vector composed of the matrices of momentum $J = 3/2$~\cite{bir_pikus,PhysRev.133.A542,luttinger_1956}. Here small cubic in $\bm J$ terms are neglected. The second contribution is a consequence of the complex valence band structure and is related to the $\bm k$-linear mixing of the heavy-hole and light-hole states by the off-diagonal elements of the Luttinger Hamiltonian $H$ and $H^*$, where $H = -\sqrt{3} \hbar^2 \gamma_3/m_0 \left( k_x - \i k_y \right) \hat{k}_z$, and $\bm k = (k_x, k_y)$ is the in-plane wave vector of the hole~\cite{kiselevmoiseev96_rus, Durnev2012797}. Here $\gamma_i$ ($i=1,2,3$) are the Luttinger parameters, $\hat{k}_z$ is the operator of a $z$-component of the hole wave vector, $\hbar$ is the Planck constant and $m_0$ is the free electron mass. The mixing results in the following form for the ground-state heavy-hole and light-hole wave functions~\cite{merkulov}
\begin{subequations}
\label{eq:wfs}
\begin{equation}
\label{eq:wflh1}
\Phi_{\pm}^{(l)} = C_{l}(z) \ket{\pm1/2} \mp \i \left(k_\mp a \right) S_{l}(z) \ket{\pm 3/2}\:,
\end{equation}
\begin{equation}
\label{eq:wfhh1}
\Phi_{\pm}^{(h)} = C_{h}(z) \ket{\pm3/2} \pm \i \left(k_\pm a \right) S_{h}(z) \ket{\pm 1/2}\:,
\end{equation}
\end{subequations}
where $C_{l}(z) \equiv \ket{lh1}$ and $C_{h}(z) \equiv \ket{hh1}$ are the envelopes of hole motion along the $z$-axis at $k = 0$, 
$\ket{\pm1/2}$, $\ket{\pm3/2}$ are the Bloch functions, $k_\pm = k_x \pm \i k_y$, and $a$ is the well width. Envelopes $C_{l,h}(z)$ satisfy the Schr\"{o}dinger equation
\begin{equation}
\left[-\frac{\hbar^2}{2m_0} \dz \left( \gamma_1 \pm 2\gamma_2 \right) \dz + V(z) \right] C_{l,h}(z) = \eps_{l,h} C_{l,h}(z)
\end{equation}
and read
\begin{equation}
\label{eq:Clh}
C_{l,h}(z) = {\cal N}_{l,h} \begin{cases} \cos k_{l,h} z,& |z| < a/2 \\ \cos \left(k_{l,h} a/2 \right) {\rm e}^{-\varkappa_{l,h} \left( |z| - a/2 \right)},& |z| > a/2 \end{cases}\:,
\end{equation}
where
\[
k_{l,h } = \sqrt{\frac{2m_0 \eps_{l,h}}{\hbar^2 (\gamma_1 \pm 2\gamma_2)}}\:, \:\:\: \varkappa_{l,h} = \sqrt{\frac{2m_0 (V_0 - \eps_{l,h})}{\hbar^2 (\gamma_1 \pm 2\gamma_2)}}\:.
\]
Here $\mathcal{N}_{l,h}$ is the normalization factor, $\eps_{l} \equiv \eps_{lh1}$ and $\eps_h \equiv \eps_{hh1}$ are the energies of $\ket{lh1}$ and $\ket{hh1}$ states, $V(z) = 0$ at $|z| < a/2$ and $V(z) = V_0$ at $|z|>a/2$ is the confinement potential of the well. The barrier height $V_0$ is equal to the valence band offset at heterointerfaces. Boundary conditions for $C_{l,h}$ and $S_{l,h}$ at the well interfaces are obtained from the continuity of $\Phi_{\pm}^{(l,h)}$ and $\hat{v}_z \Phi_{\pm}^{(l,h)}$ columns with $\hat{v}_z$ being the velocity operator. 
Neglecting the terms of the second order in $ka$ these boundary conditions for $C_{l,h}$ coincide with the ones proposed by Bastard~\cite{PhysRevB.24.5693}.

Envelopes $S_{l,h}(z)$ are odd in $z$ and satisfy the following equation
\begin{multline}
\label{eq:S}
\left[ -\frac{\hbar^2}{2m_0} \dz \left( \gamma_1 \mp 2\gamma_2 \right) \dz + V(z) - \eps_{l,h} \right] S_{l,h}(z) = \\
= -\frac{\sqrt{3} \hbar^2}{m_0 a} \left \{ \gamma_3 \dz \right \} C_{l,h}(z)\:.
\end{multline}
Here the upper and lower signs correspond to $S_l$ and $S_h$, respectively, curly brackets define the symmetrized product $\left \{ \gamma_3 \dz \right \} = \frac12 \left( \gamma_3 \dz + \dz \gamma_3 \right)$.
With account for Eq.~\eqref{eq:Clh} the solutions of Eq.~\eqref{eq:S} can be written as
\begin{subequations}
\label{eq:Slh}
\begin{equation}
S_l = \begin{cases} A_1^{(l)} \sin \left( k_l z/\sqrt{\nu} \right) + A_2^{(l)} \sin k_l z,& |z| < a/2 \\ 
\left[ B_1^{(l)} {\rm e}^{-\varkappa_l \left( |z| - a/2 \right)/\sqrt{\nu}} + \right.  \\ \left. + B_2^{(l)} \cos \left(k_{l} a/2 \right) {\rm e}^{-\varkappa_{l} \left( |z| - a/2 \right)} \right] {\rm sign} (z),& |z| > a/2 \end{cases}\:,
\end{equation}
\begin{equation}
S_h = \begin{cases} A_1^{(h)} \sin \left( k_h \sqrt{\nu} z \right) + A_2^{(h)} \sin k_h z,& |z| < a/2 \\ 
\left[ B_1^{(h)} {\rm e}^{-\varkappa_h \sqrt{\nu} \left( |z| - a/2 \right)} + \right. \\ \left. + B_2^{(h)} \cos \left(k_{h} a/2 \right) {\rm e}^{-\varkappa_{h} \left( |z| - a/2 \right)} \right] {\rm sign}(z),& |z| > a/2 \end{cases}\:,
\end{equation}
\end{subequations}
where $\nu = (\gamma_1 - 2\gamma_2)/ (\gamma_1 + 2\gamma_2)$.
Coefficients $A_2^{(l,h)}$ and $B_2^{(l,h)}$ are found directly from Eq.~\eqref{eq:S} and read
\begin{equation}
A_2^{(l,h)} = -\frac{\sqrt{3}}{2 k_{l,h} a} \frac{\gamma_3}{\gamma_2} \mathcal{N}_{l,h}\:,\:\:\:B_2^{(l,h)} = \frac{\sqrt{3}}{2 \varkappa_{l,h} a} \frac{\gamma_3}{\gamma_2} \mathcal{N}_{l,h} \:,
\end{equation}
while to derive $A_1^{(l,h)}$ and $B_1^{(l,h)}$ one should employ the boundary conditions for $S_{l,h}$. Continuity of $\hat{v}_z \Phi_{\pm}^{(l,h)}$ results in 
\begin{equation}
\left. \left[ \left( \gamma_1 \mp 2\gamma_2 \right) \dz S_{l,h} + \frac{\sqrt{3}}{a} \gamma_3 C_{l,h} \right] \right \vert_{z_i -}^{z_i+} = 0\:,
\end{equation}
where $z_i$ is the interface coordinate. The same relation is obtained after integration of Eq.~\eqref{eq:S} over the interface.

In the presence of magnetic field the cycle components of the wave vector $k_\pm$ which enter the $H$ and $H^*$ operators should be written as $k_\pm - |e|/(c\hbar) A_\pm$, where $A_\pm = A_x \pm \i A_y$, $\bm{A}$ is the vector potential of the field, $e$ is the electron charge and $c$ is the speed of light. It results in a linear in magnetic field correction to the energies of $\Phi_\pm^{(l,h)}$ states described by the following $g$-factors  
\begin{subequations}
\label{eq:gfactor_all}
\begin{equation}
\label{eq:glh1_all}
g_{lh1} = -2\varkappa + 4\sqrt{3} a \braket{S_l(z)}{\left \{ \gamma_3 \dz \right \}}{C_l(z)}\:,
\end{equation}
\begin{equation}
\label{eq:ghh1_all}
g_{hh1} = -6\varkappa + 4\sqrt{3} a \braket{S_h(z)}{\left \{ \gamma_3 \dz \right \}}{C_h(z)}\:.
\end{equation}
\end{subequations}
Hereafter angular brackets denote the quantum-mechanical average. It is noteworthy that matrix elements in the angular brackets coincide with the sums of the following series derived in the framework of perturbation theory
(cf.~[\onlinecite{PhysRevB.50.8889}] and [\onlinecite{Durnev2012797}])
\begin{subequations} 
\label{eq:gfactor}
\begin{multline}
\label{eq:glhn}
\braket{S_l(z)}{\left \{ \gamma_3 \dz \right \}}{C_l(z)} = \\
= \frac{\sqrt{3} \hbar^2}{m_0 a} \sum \limits_\nu \frac{\left|\left\langle
      hh,\nu|\left \{ \gamma_3\hat{k}_{z} \right \}|lh1\right\rangle \right|^2}{\eps_{lh1} - \eps_{hh,\nu}}\:, 
\end{multline}
\begin{multline}
\label{eq:ghhn}
\braket{S_h(z)}{\left \{ \gamma_3 \dz \right \}}{C_h(z)} = \\
= \frac{\sqrt{3} \hbar^2}{m_0 a} \sum \limits_\nu \frac{\left|\left\langle
      hh1|\left\{ \gamma_3\hat{k}_{z} \right \}|lh, \nu\right\rangle \right|^2}{\eps_{lh, \nu} - \eps_{hh1}}\:,
\end{multline}
\end{subequations}
where index $\nu$ enumerates the states of both discrete and continuum spectra. Hereafter we will use the hole representation which corresponds to positive values of the size-quantization energies $\eps_{lh,\nu}$ and $\eps_{hh,\nu}$. In the considered case of a symmetric rectangular well non-zero matrix elements of $\{\gamma_3\hat{k}_{z}\}$ entering the sums, Eq.~\eqref{eq:gfactor}, exist for even $\nu$ only.

Typically the values of $\eps_{lh1}$ and $\eps_{hh2}$ are close in quantum wells giving rise to a ``resonant'' contribution in the sum for $g_{lh1}$~\cite{Durnev2012797}. On the contrary, the sum for $g_{hh1}$ does not contain such a contribution which explains larger values of  $g_{lh1}$ observed in experiments. To illustrate this let us calculate the values of $g_{lh1}$ and $g_{hh1}$ for the well with infinitely high barriers. For the GaAs parameters (see parametrization (A) in Tab.~\ref{table}) the renormalization of the $g$-factor with respect to its bulk value $\Delta g_{lh1} = g_{lh1} + 2\varkappa \approx 22.4$ which is considerably larger than $\Delta g_{hh1} = g_{hh1} + 6\varkappa \approx 2.6$. 

\subsection{\label{sec:nonlinear}Account for the interface mixing}
An important effect that influences the values of $g_{lh1}$ in quantum wells is the interface mixing of heavy and light holes~\cite{aleiner, ivchenko96}. This mixing is described by the following Hamiltonian
\begin{equation}
{\cal H}_{l-h} = \pm t_{l-h} \left( \hbar^2/\sqrt{3}m_0a_0 \right)\left\{J_x J_y \right\} \delta(z-z_i)
\end{equation}
with the dimensionless parameter $t_{l-h}$ (on the order of 1 in GaAs/AlGaAs quantum wells~[\onlinecite{ivchenko96}]). Here the signs $+$ and $-$ correspond to the left and the right well interfaces, curly brackets denote the symmetrized product of operators. Hereafter we assume the proximity of $lh1$ and $hh2$ levels which allows one to use the resonant approximation considering $\ket{lh1, \pm 1/2}$ and $\ket{hh2, \mp3/2}$ states only. As a result of the interface mixing the hole state in a quantum well is described at $k = 0$ by the wave functions $\Psi_{\pm}^{(j)}$ ($j = \pm 1$ is the spin index) being linear combinations of $\ket{lh1, \pm 1/2}$ and $\ket{hh2, \mp3/2}$ with coefficients $\mathcal C_l$ and $\mathcal C_h$~\cite{Durnev2012797}
\begin{eqnarray}
\label{eq:wfsint}
\Psi_{-}^{(j)} &=& \mathcal{C}_l \ket{lh1, 1/2j} +  \mathrm{i} j \mathcal{C}_{h} \ket{hh2,-3/2j} \nonumber \\
\Psi_{+}^{(j)} &=& \mathcal{C}_h \ket{lh1, 1/2 j} - \mathrm{i} j \mathcal{C}_{l} \ket{hh2,-3/2j}\:.
\end{eqnarray}
$\Psi_\pm^{(j)}$ states are spin-degenerate at zero magnetic field having the energies $\eps_+ \equiv \eps_+^{(j)}$ and $\eps_- \equiv \eps_-^{(j)}$ (hereafter we assume $\eps_+ > \eps_-$). 
 
\begin{figure}[hptb]
\includegraphics[width=0.47\textwidth]{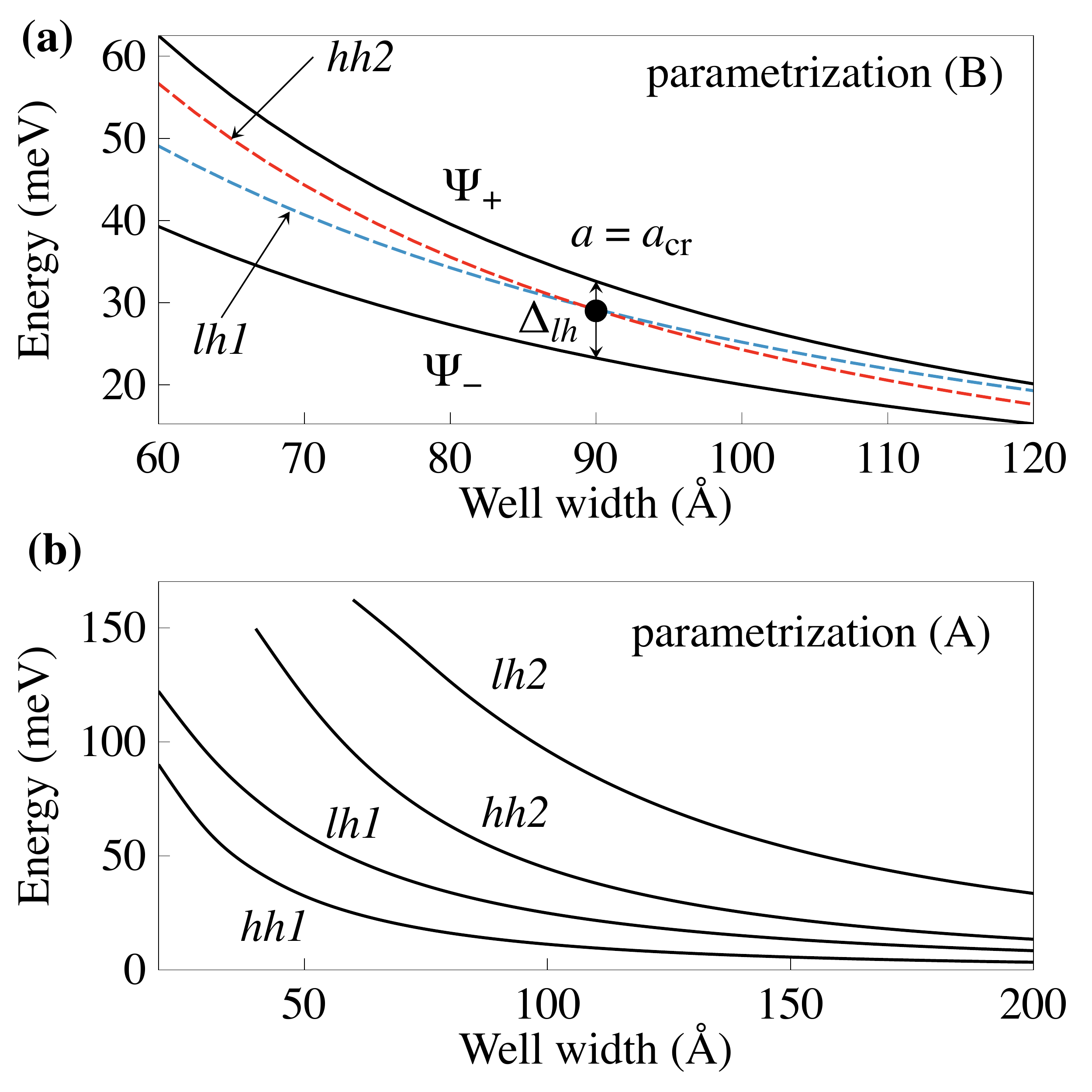}
\caption{\label{fig:fig3} Energy spectrum of holes at $k=0$ in the GaAs/Al$_{0.35}$Ga$_{0.65}$As quantum well. (a) Calculations for the parametrization (B): $lh1$ and $hh2$ levels (dashed lines) and the energies of the mixed states $\Psi_\pm$ at $t_{l-h} = 1$ (solid curves) are shown. (b) Calculations for the set of parameters (A).}
\end{figure}

Figure~\ref{fig:fig3}a shows $\eps_+$ and $\eps_-$ dependences on the GaAs/Al$_{0.35}$Ga$_{0.65}$As well width. Allowance for the interface mixing results in an anti crossing of the hole levels (solid and dashed lines in Fig.~\ref{fig:fig3}a) with the splitting $\Delta_{lh} = 2|\braket{lh1,\pm1/2}{{\cal H}_{l-h}}{hh2,\mp3/2}|$ in the critical point $a = a_{cr}$ (well width which corresponds to $\eps_{lh1} = \eps_{hh2}$). Interface effects also considerably influence the in-plane energy dispersion of $\Psi_{\pm}^{(j)}$ subbands (see Refs.~\cite{durnev2, toropov} for details).

In the resonant approximation Zeeman splitting of $\Psi_+$ and $\Psi_-$ states in the presence of longitudinal magnetic field $B_z$ can be obtained via diagonalization of the effective 4$\times$4 Hamiltonian written in the basis of vectors $\left( a_n \ket{n}, b_{n-1} \ket{n-1}, c_n \ket{n}, d_{n-1} \ket{n-1}\right)^{T}$ comprising Landau levels $\ket{n}$, $n=0,1..$, (cf.~\cite{rashba60}):
\begin{equation}
\label{eq:HamB}
{\cal H}_B = \left( 
\begin{array}{cccc}
\eps_{hh2} & -\i \alpha \sqrt{n B_z} & \i \Delta_{lh}/2 & 0\\
\i \alpha \sqrt{n B_z} & \eps_{lh1} & 0 & \i \Delta_{lh}/2\\
-\i \Delta_{lh}/2 & 0 & \eps_{lh1} & -\i \alpha \sqrt{n B_z}\\
0 & -\i \Delta_{lh}/2 & \i \alpha \sqrt{n B_z} & \eps_{hh2}\\
\end{array}
\right)\:,
\end{equation}
\[
\alpha = 2\sqrt{3}\sqrt{\mu_B \hbar^2/m_0} \left|\left\langle hh2|\left \{ \gamma_3\hat{k}_{z} \right \}|lh1\right\rangle \right|\:.
\]
We will be further interested only in the first two terms in $\sqrt{B_z}$ in the expansion of Zeeman splitting, therefore in Eq.~\eqref{eq:HamB} we neglect diagonal cyclotron energies which lead to the contributions $\propto B_z^{3/2}$ and $\propto B_z^2$. We also disregard a linear in $B_z$ ``bulk'' contribution~\eqref{eq:bulk} since it is relatively small comparing to the renormalization induced by the mixing of valence subbands. The eigenenergies of Eq.~\eqref{eq:HamB} at $n = 0$ equal to $\eps_+$ and $\eps_-$ and correspond to the energies of $\Psi_+^{(j)}$ and $\Psi_-^{(j)}$ at zero magnetic field. The spin-degeneracy is lifted at $n>0$ resulting in two pairs of levels with energies
\begin{eqnarray}
\eps_+^{(j)}(n) &=& \frac12 \sqrt{\left( \eps_{lh1} - \eps_{hh2} \right)^2 + \left( \Delta_{lh} + 2 j \alpha \sqrt{n B_z} \right)^2} \nonumber \\
\eps_-^{(j)}(n) &=& -\frac12 \sqrt{\left( \eps_{lh1} - \eps_{hh2} \right)^2 + \left( \Delta_{lh} - 2 j \alpha \sqrt{n B_z} \right)^2}\:. \nonumber \\
\end{eqnarray}
We will define the Zeeman splitting of the $n$-th Landau level as $\Delta E_{z, \pm} (n) = \eps_\pm^{(+1)} (n+1) - \eps_\pm^{(-1)} (n)$ yielding for $n=0$
 \begin{multline}
\label{eq:Zeem}
\Delta E_{z, \pm} = \left| \frac12 \sqrt{(\eps_{lh1} - \eps_{hh2})^2 + \left( \Delta_{lh} \pm 2\alpha \sqrt{B_z} \right)^2} - \right.\\ 
\left.- \frac12 \sqrt{(\eps_{lh1} - \eps_{hh2})^2 + \Delta_{lh}^2} \right|\:.
\end{multline}
with $\Delta E_{z, \pm} \equiv \Delta E_{z, \pm} (0)$.
As it is seen from Eq.~\eqref{eq:Zeem}, the behavior of the splitting with magnetic field depends on the relation between the energy parameters $\Delta_{lh}$ and $\eps_{lh1}-\eps_{hh2}$. Particularly, in the critical point one has $\Delta E_{z,\pm}\propto \sqrt{B_z}$. In the opposite limit, $\Delta_{lh} \to 0$, $\eps_{lh1} - \eps_{hh2} \neq 0$, the Zeeman splitting in small fields is linear in $B_z$ and is described by the $g$-factor, Eq.~\eqref{eq:glh1_all}, \eqref{eq:glhn}, calculated in the resonant approximation (in the sum Eq.~\eqref{eq:glhn} one should leave $\nu = 2$ only).
In the case when $\Delta_{lh}$ and $\eps_{lh1}-\eps_{hh2}$ are comparable Zeeman splitting in small fields is given by
\begin{equation}
\label{eq:Zeem_approx}
\Delta E_{z,\pm}  \approx \frac{\alpha\Delta_{lh}}{\Delta} \sqrt{B_z} + \frac{\alpha^2}{\Delta} B_z\:,
\end{equation}
\[
\Delta = \eps_+ - \eps_- = \sqrt{(\eps_{lh1} - \eps_{hh2})^2 + \Delta_{lh}^2}
\]
and comprises the sum of linear and square-root contributions. Nonlinear contribution in Eq.~\eqref{eq:Zeem_approx} is related to appearance of the interface-induced $\bm k$-linear terms in the effective Hamiltonian of $\Psi_{\pm}^{(j)}$ states (see Ref.~\cite{durnev2} for details). The presence of such terms leads to the square-root dependence of Landau levels on magnetic field~\cite{rashba60}.
The transition between two regimes occurs at a critical field $B^* = \Delta_{lh}^2/\alpha^2$, therefore the nonlinear region increases with the increase of $t_{l-h}$. Considering the well presented in Fig.~\ref{fig:fig3}a at $a = 70$~\AA~and $t_{l-h} = 1$ one has $\eps_{lh1} - \eps_{hh2} \approx -3.6$~meV, $\Delta_{lh} \approx 15$~meV and $B^* \approx 11$~T. However it is worth to mention that Zeeman splitting in such field $\left| \Delta E_{z,\pm} (B^*) \right| \approx 15$~meV and can be comparable with the distance to other size-quantized levels, therefore the resonant model is no longer applicable. Square-root term in Eq.~\eqref{eq:Zeem_approx} restricts the linear approximation of the Zeeman splitting with the effective $g$-factor $g_{\pm} = \alpha^2/(\Delta \mu_B)$ to a particular range of magnetic fields $B_z > B^*$.

\subsection{Role of excitonic effects}
The above presented theory for Zeeman effect is valid for holes which move freely in a quantum well plane. However all the experimental data listed in Tab.~\ref{table1} were obtained by means of optical spectroscopy in the region of excitonic transitions, thus Coulomb interaction between an electron and hole might play an important role~\cite{toropov}. In our examination we will restrict ourselves to the limit of a strong confinement along the $z$-axis (the Bohr radius of a three-dimensional exciton is larger than the well width, $a_B > a$) and the resonant approximation. Therefore the wave function of an exciton comprising an electron in the ground state ($e1$) and a hole reads
\begin{equation}
\label{eq:wf_exc}
\ket{X_{\pm,l}(s;j)} = \frac{{\rm e}^{\mathrm{i} \bm K \bm R}}{\sqrt{S}} \psi_l (\bm r) \ket{e1,1/2s} \Psi_\pm^{(j)}\:.
\end{equation}
Here $\bm K$ and $\bm R$ are the wave vector and the coordinate of an exciton center of mass, $S$ is the normalization area, $\ket{e1,1/2s}$ is the wave function of the electron, $s = \pm 1$ denotes its spin index, $\psi_l (\bm r)$ is the wave function of electron-hole relative motion in the well plane, index $l = 1s, 2s, 2p \dots$ numerates the states of the relative motion.
We will further use the model of a two-dimensional exciton which allows one to obtain analytical expressions for $\psi_l (\bm r)$ and the exciton energy spectrum. Precise calculation of $\psi_l (\bm r)$ requires the account for large $\bm k$-linear terms in the energy spectrum of hole subbands which is beyond the scope of the present work~\cite{kovalev}.
The wave functions of hole motion along the $z$-axis including the interface effects, $\Psi_\pm^{(j)}$, are defined by Eq.~\eqref{eq:wfsint}. In what follows we will restrict our analysis to the states with lower energy described by the $\ket{X_{-,l}(s;j)}$ functions. These states are  optically active in the case of coinciding spin indices of an electron and hole, i.e. at $s = j = 1$ and $s = j = -1$. 

Magnetic field affects only the electron-hole relative motion which is quantized in the presence of Coulomb interaction. This defines the linear behavior of Zeeman splitting with magnetic field even in the presence of the interface mixing of holes. Zeeman splitting is therefore described by the $g$-factor
\begin{equation}
g(X_{-,1s}) = g^{(e)}(X_{-,1s}) + g^{(h)}(X_{-,1s})\:,
\end{equation}
which consists of an electronic and hole contributions. The electronic component of the $g$-factor is defined mainly  by the width and composition of the well~\cite{ivchenko_kiselev92, PhysRevB.75.245302}, while its renormalization due to Coulomb effects is negligible. 

Renormalization of the hole component is analogous to that of a free hole and is controlled by the mixing of $X_{-,1s}$ and $X_{\pm,\nu p}$ ($\nu = 2,3\dots$) states in the framework of the Luttinger Hamiltonian. Since these states are separated in energy, the hole contribution to the excitonic $g$-factor can be presented as a sum of the following perturbation series
\begin{multline}
\label{eq:exc1}
g^{(h)}(X_{-,1s}) = -2\varkappa\left( \left| \mathcal{C}_l \right|^2 - 3 \left| \mathcal{C}_h \right|^2 \right) - 12 \frac{\hbar^2}{m_0 a^2} \zeta^2 C_x \times \\
\times \sum_\nu \left[4\left| \mathcal{C}_l \right|^2 \left| \mathcal{C}_h \right|^2 \frac{\braket{\psi_{1s}}{k_-}{\psi_{\nu p}} \braket{\psi_{\nu p}}{r_+}{\psi_{1s}}}{E(X^{-}_{1s}) - E(X^{-}_{\nu p})} \right. \\
\left. + \left(\left| \mathcal{C}_l \right|^2 - \left| \mathcal{C}_h \right|^2 \right)^2 \frac{\braket{\psi_{1s}}{k_-}{\psi_{\nu p}} \braket{\psi_{\nu p}}{r_+}{\psi_{1s}}}{E(X^{-}_{1s}) - E(X^{+}_{\nu p})} \right]\:.
\end{multline}
Here we introduced the parameter $\zeta = a\left|\left\langle hh2|\left \{ \gamma_3\hat{k}_{z} \right \}|lh1\right\rangle \right|$, while $k_\pm = k_x \pm \mathrm{i} k_y$ and $r_\pm = x \pm \mathrm{i} y$ are the cycle components of the wave vector and coordinate of the electron-hole relative motion.  Coefficient $C_x \sim 1$ is determined by the dispersion of an electron-hole pair and the exciton structure in the vicinity of $k = 0$. Let us stress that the summation by $\nu$ is performed over the states of both discrete and continuum spectra of the Coulomb problem. Equation~\eqref{eq:exc1} can be simplified using the relation $\braket{\psi_{1s}}{k_-}{\psi_{\nu p}} = (\mathrm{i} \mu/\hbar^2) \braket{\psi_{1s}}{r_-}{\psi_{\nu p}} \left(E_{1s} - E_{\nu p} \right)$, where $\mu$ is the reduced mass of the exciton, $E_{1s} \equiv E(X_{-,1s})$ and $E_{\nu p} \equiv E(X_{-,\nu p})$, and completeness of the $\psi_l$ set. After simplifications $g^{(h)}(X_{-,1s})$ can be written in the form
\begin{multline}
\label{eq:exc2}
g^{(h)}(X_{-,1s}) = -2\varkappa\left( \left| \mathcal{C}_l \right|^2 - 3 \left| \mathcal{C}_h \right|^2 \right) - 24 \zeta^2 C_x \frac{\hbar^2}{m_0 a^2 E_{1s}}  \times \\
\times \left[ 4\left| \mathcal{C}_l \right|^2 \left| \mathcal{C}_h \right|^2 \left| \braket{\psi_{1s}}{\bar{r}^2}{\psi_{1s}} \right| \right. \\
\left. + \left(\left| \mathcal{C}_l \right|^2 - \left| \mathcal{C}_h \right|^2 \right)^2 \sum_\nu \frac{E_{1s} - E_{\nu p}}{E_{1s} - E_{\nu p} - \Delta} \left| \braket{\psi_{1s}}{\bar{r}_-}{\psi_{\nu p}} \right|^2 \right]\:,
\end{multline}
where we used the relation $E_{1s} = 2\hbar^2/\mu a_B^2$, and introduced the operators of the dimensionless coordinates $\bar{r}^2 = r^2/a_B^2$ and $\bar{r}_- = r_-/a_B$. 

Equation~\eqref{eq:exc2} represents the light-hole $g$-factor with the account for Coulomb effects. 
 In the case of $\Delta_{lh} = 0$ and $E_{1s} \ll \Delta$ (the limit corresponding to a free hole), Eq.~\eqref{eq:exc2} yields the formula Eq.~\eqref{eq:glh1_all}, \eqref{eq:glhn} for $g_{lh1}$ in the resonant approximation ($\nu = 2$). In the critical point at $\Delta_{lh} \neq 0$ one has $\left| \mathcal{C}_l \right|^2 = \left| \mathcal{C}_h \right|^2 = 1/2$, and the second term in Eq.~\eqref{eq:exc2} vanishes yielding 
\begin{equation}
\label{eq:gcritical}
g^{(h)}(X_{-,1s}) = 2\varkappa - \frac92 \zeta^2 C_x \frac{\mu}{m_0} \left( \frac{a_{B}}{a} \right)^2\:.
\end{equation}
For a GaAs/Al$_{0.35}$Ga$_{0.65}$As quantum well at $a = a_{cr}$ (see Fig.~\ref{fig:fig3}a), $C_x = 1$ and $E_{1s} = 10$~meV the second term in Eq.~\eqref{eq:gcritical} is estimated as $\approx - 21$.

\section{Results and discussions}

Below we will analyze results of calculations of Zeeman splitting for heavy and light holes in different approaches discussed in Sec.~II.
\subsection{Heavy-hole $g$-factor}

\begin{figure}[hptb]
\includegraphics[width=0.47\textwidth]{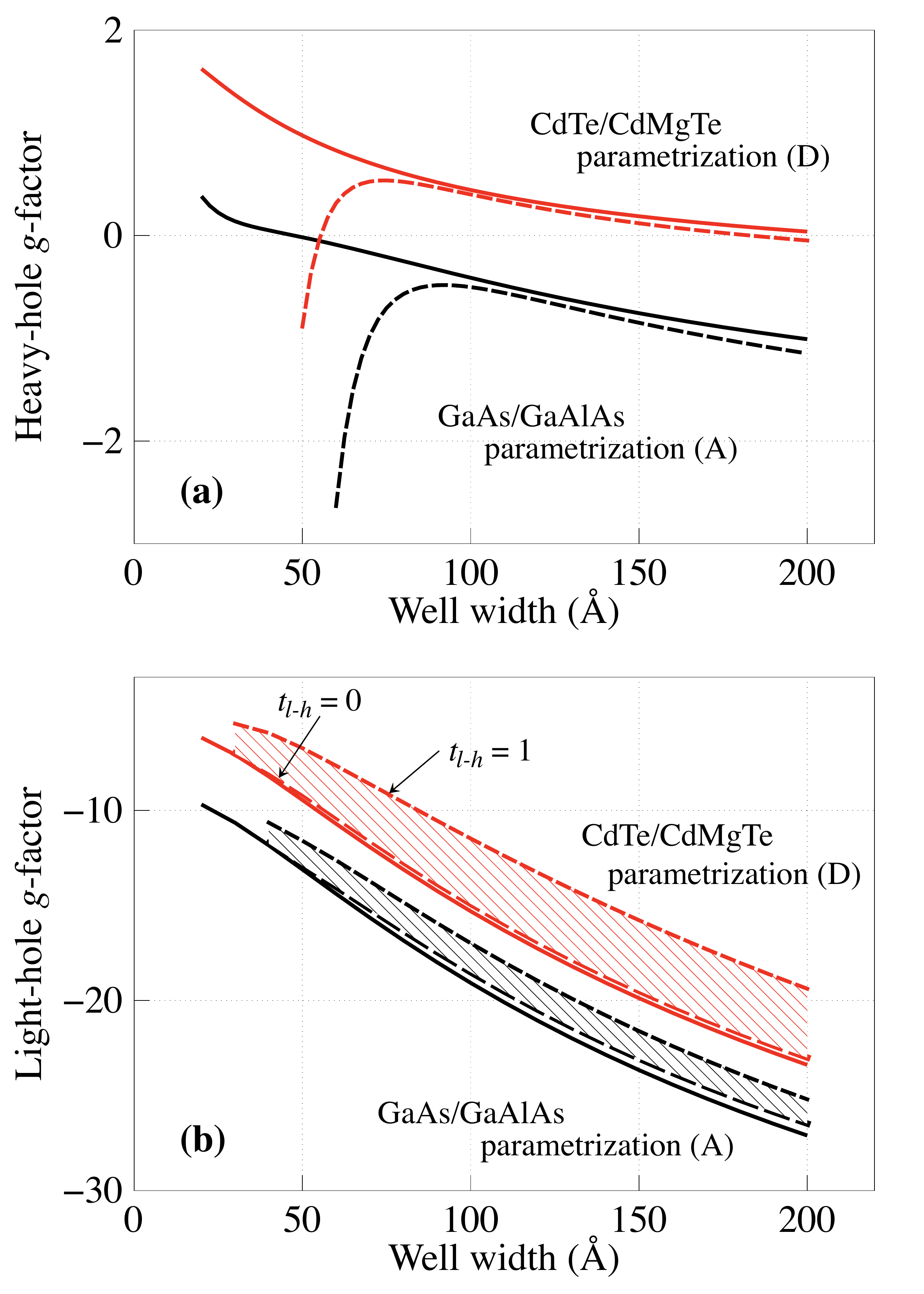}
\caption{\label{fig:fig2} The calculated heavy-hole (a) and light-hole (b) $g$-factors as a function of the well width for two quantum wells, GaAs/Al$_{0.35}$Ga$_{0.65}$As and CdTe/Cd$_{0.74}$Mg$_{0.26}$Te. Solid curves correspond to the multi-level approximation Eq.~\eqref{eq:gfactor_all}. Dashed curves stand for the calculations in the resonant approach at $t_{l-h} = 0$ (a) and $t_{l-h} = 0$ and 1 (b).}
\end{figure}

Let us first consider a free hole at $t_{l-h} = 0$. 
The results of calculations of the heavy-hole $g$-factor $g_{hh1}$ following Eq.~\eqref{eq:ghh1_all} are shown in Fig.~\ref{fig:fig2}a by the solid lines.
The calculations were performed for two systems, namely GaAs/Al$_{0.35}$Ga$_{0.65}$As and CdTe/Cd$_{0.74}$Mg$_{0.26}$Te, with the use of parameters listed in Tab.~\ref{table}. 
For GaAs-based structures the results are presented for two parametrizations of the Luttinger hamiltonian [parametrizations (A) and (B) in Tab.~\ref{table}] resulting in similar energy spectra of the heavy hole $hh1$ (see also Ref.~\cite{durnev2}). For comparison the results of calculations in the resonant approximation are presented in Fig.~\ref{fig:fig2}a by the dashed lines. As $\eps_{lh,\nu} - \eps_{hh1} > 0$ for any value of $\nu$ in unstrained zinc-blende-based wells, the renormalization of $g_{hh1}$ in such structures is positive [see Eq.~\eqref{eq:ghhn}]. Figure~\ref{fig:fig2}a demonstrates that the most striking difference between the multi-level and the resonant approaches is observed for $g_{hh1}$ in the region of thin wells ($a < 70$~\AA) owing to the shift of the $lh2$ level towards a continuum spectrum. The discrepancy of two approaches in thick wells is related to the localization of the excited states with large values of $\nu$ which are disregarded in the resonant model. The calculations based on Eq.~\eqref{eq:ghh1_all} in the multi-level approach are in a quantitive agreement with the calculations for the same systems performed in Ref.~[\onlinecite{kiselevmoiseev96_rus}] by means of the numeric diagonalization of the 8-band Hamiltonian in the presence of magnetic field. Moreover Fig.~\ref{fig:fig2}a demonstrates the change of the $g_{hh1}$ sign which qualitatively agrees with the experimental data (see Ref.~\cite{PhysRevB.45.3922} for example), though a particular value of the well width corresponding to the sign change is governed by the parametrization used. 

\begin{table}
\caption{Different parametrizations of the Luttinger Hamiltonian used in the $g$-factor calculations. The values of $\gamma_i$ for parametrizations (A) and (B) are calculated with the use of 14-band parameters. The values of $\gamma_i$, $\varkappa$ and energy gap of the solid solutions follow linear and parabolic interpolations, respectively. The parameters of  CdMgTe are chosen to be the same as of CdTe.}
 \begin{tabular}{p{0.1\linewidth} p{0.3\linewidth} p{0.1\linewidth} p{0.1\linewidth} p{0.1\linewidth} p{0.1\linewidth} c}
\hline
\hline
 & Material & $\gamma_1$ & $\gamma_2$ & $\gamma_3$ & $\Delta E_v$ (meV) & $\varkappa$ \\
\hline
А \cite{jancu:193201} & GaAs & 8.5 & 2.08 & 3.53 & -- & 1.2\\
А \cite{jancu:193201} & Al$_{0.35}$Ga$_{0.65}$As  & 7.17 & 1.52 & 2.9 & 168 & 0.82\\
\hline
B \cite{maruschak1988} & GaAs & 7.5 & 2.7 & 3.4 & -- & 1.2\\
B \cite{maruschak1988} & Al$_{0.35}$Ga$_{0.65}$As & 6.19 & 2.04 & 2.71 & 168 & 0.82\\
\hline
C \cite{vurgaftman02} & In$_{0.53}$Ga$_{0.47}$As  & 13.7 & 5.95 & 5.95 & 354 & 4.63\\
C \cite{vurgaftman02} & InP & 5& 1.84 & 1.84 & -- & 0.97 \\
\hline
D \cite{PSSB:PSSB2221250242} & CdTe & 4.72& 1.29 & 1.85 & -- & 0.47 \\
D \cite{PSSB:PSSB2221250242} & Cd$_{0.74}$Mg$_{0.26}$Te & 4.72 & 1.29 & 1.85 & 133 & 0.47 \\
\hline
\hline
\end{tabular}
\label{table}
\end{table}

\subsection{Nonlinear Zeeman effect and the light-hole $g$-factor}

The role of parametrization appears to be particularly important in the calculations of the light-hole $g$-factor (Fig.~\ref{fig:fig2}b). This is related to the fact that the renormalization of $g_{lh1}$ can be either positive or negative depending on the sign of a ``resonant'' energy denominator $\eps_{lh1} - \eps_{hh2}$. For illustration positions of hole size-quantized levels are shown in Fig.~\ref{fig:fig3}. One can see that for parametrization (A) $lh1$ level has lower energy than $hh2$ for the whole range of well width (Fig.~\ref{fig:fig3}b), while for parametrization (B) (Fig.~\ref{fig:fig3}a) these levels cross each other at the critical width $a_{cr} \approx 90$~\AA. In this point the value of $g_{lh1}$ calculated following Eq.~\eqref{eq:glh1_all} goes to infinity. The dashed lines in Fig.~\ref{fig:fig2}b represent the results of calculations of the effective light-hole $g$-factor corresponding to the linear in magnetic field term in Eq.~\eqref{eq:Zeem_approx}, which were performed following Eq.~(9a) of Ref.~\cite{Durnev2012797}. The calculations are performed for the quantum wells GaAs/Al$_{0.35}$Ga$_{0.65}$As and CdTe/Cd$_{0.74}$Mg$_{0.26}$Te for $t_{l-h}$ lying in the range from 0 to 1. It is seen that the allowance for the interface mixing results in the considerable decrease of the absolute value of the $g$-factor due to the increase of $\eps_+ - \eps_-$.

\begin{figure}[hptb]
\includegraphics[width=0.47\textwidth]{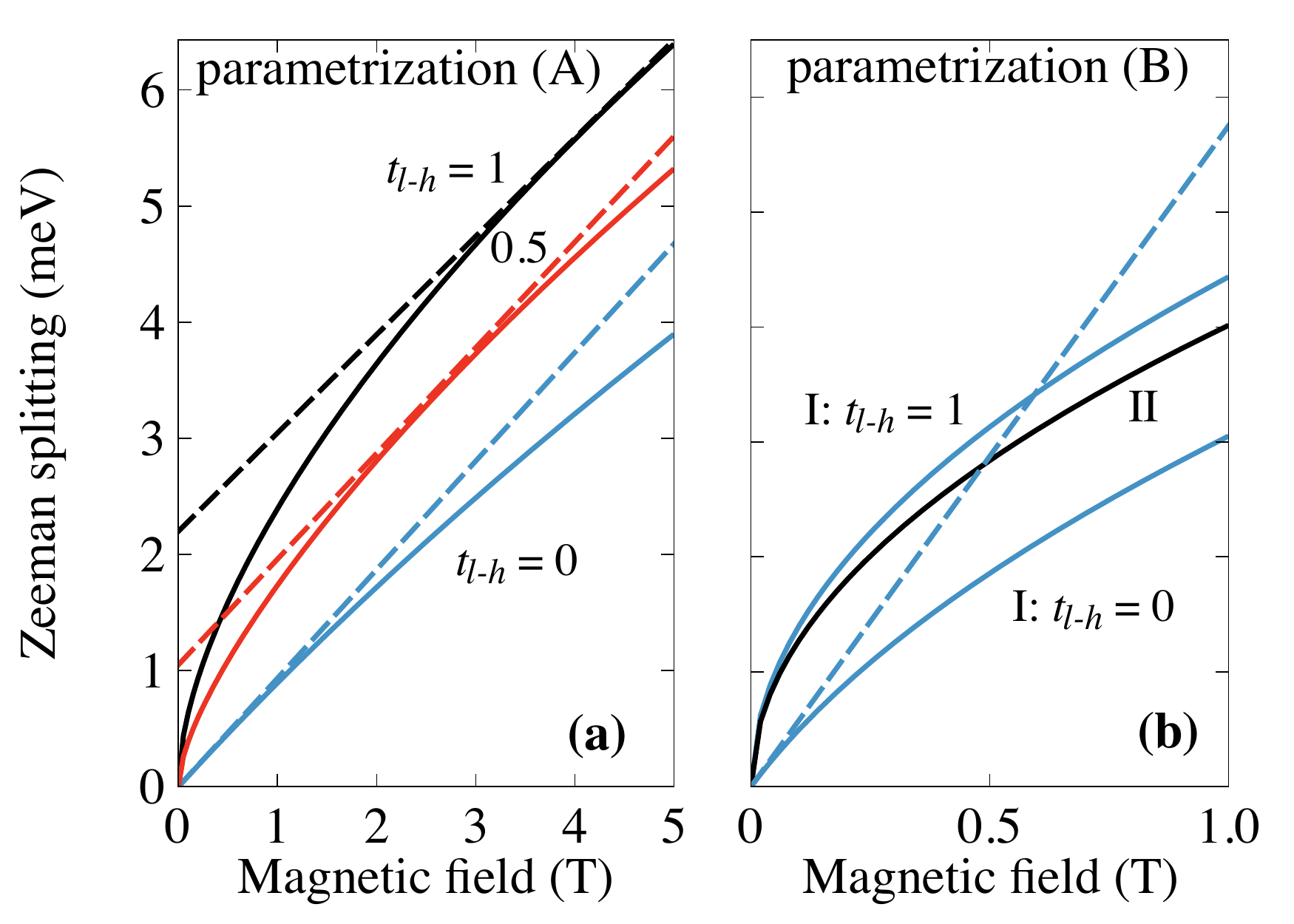}
\caption{\label{fig:fig4} Zeeman splitting of the $\Psi_{+}$ state calculated following Eq.~\eqref{eq:Zeem}. (a) Calculations for the set of parameters (А) at $a = 100$~\AA. The values of the critical fields $B^* \approx 1.1$~T ($t_{l-h} = 0.5$) and $B^* \approx 4.6$ T ($t_{l-h} = 1$). (b) Calculations for the set of parameters (B): $a = 70$~\AA~$< a_{cr}$ (curve I), $a=a_{cr}$ (curve II). The value of the critical field for the curve I ($t_{l-h} = 1$) $B^* \approx 11$~T. 
Dashed curves represent the linear approximation by the second term of Eq.~\eqref{eq:Zeem_approx}.}
\end{figure}

As it was mentioned in Sec.~\ref{sec:nonlinear}, Zeeman splitting of light hole at $t_{l-h} \neq 0$ is given by the sum of linear and square-root contributions [see Eq.~\eqref{eq:Zeem_approx}]. It means that the linear approximation based on the effective $g$-factor presented in Fig.~\ref{fig:fig2}b is valid in a particular range of magnetic field $B_z > B^*$ only. Fig.~\ref{fig:fig4} shows Zeeman splitting calculated following Eq.~\eqref{eq:Zeem_approx}. It is seen that the increase of $t_{l-h}$ is accompanied by the increase of $B^*$ (see Fig.~\ref{fig:fig4}a) leading to the increase of the region where the linear approximation is inapplicable. In the case of parametrization (B) and in the presence of the interface mixing ($t_{l-h} = 1$) in the whole reasonable range of magnetic fields accessed in experiments one has $\Delta E_{z,\pm} \propto \sqrt{B_z}$ (Fig.~\ref{fig:fig4}b), and the linear approximation fails. 

\section{Comparison of theory and experiments}

\begin{figure}[hptb]
\includegraphics[width=0.49\textwidth]{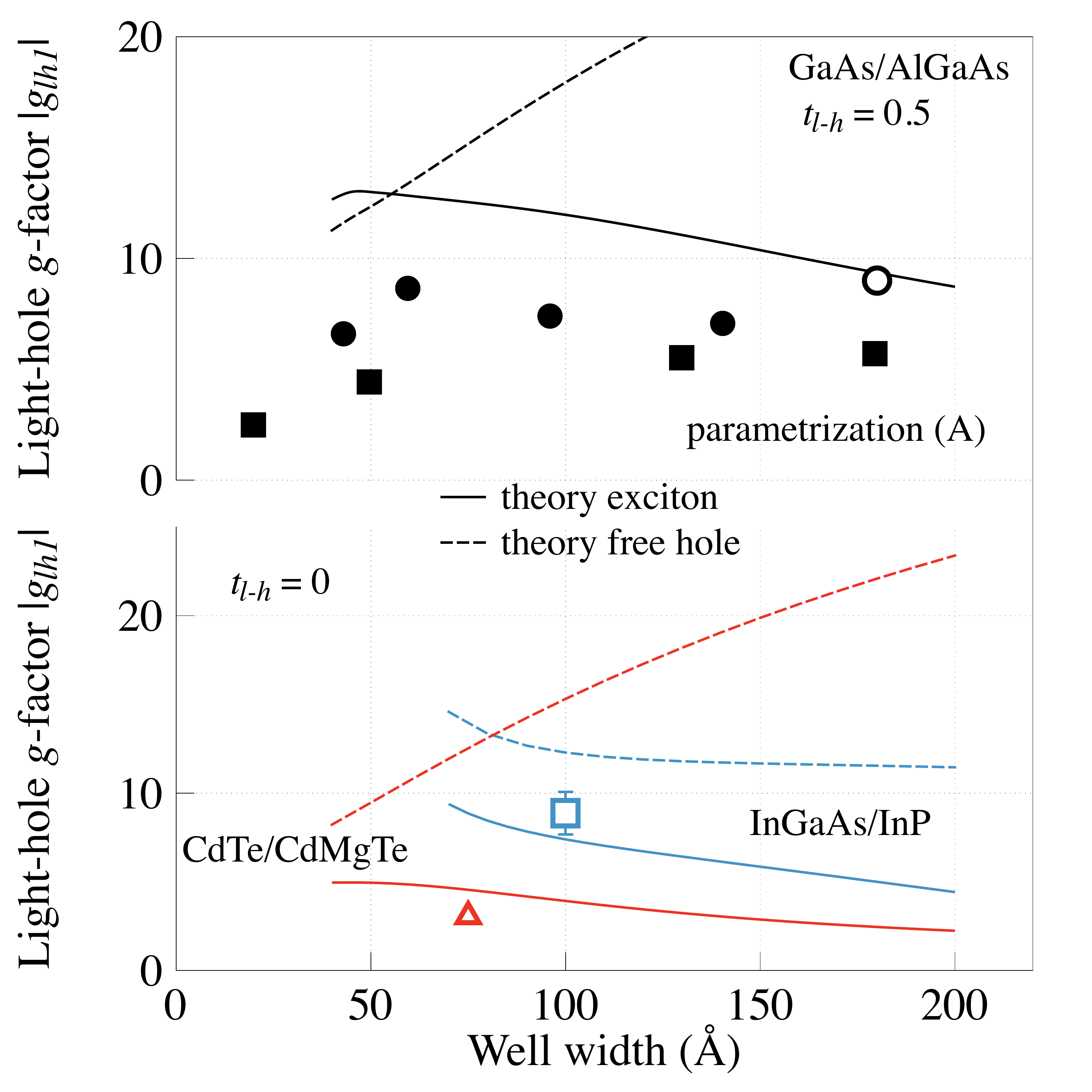}
\caption{\label{fig:fig1} The light-hole $g$-factor: comparison of theory and experiments. The dots represent the data of Ref.~[\onlinecite{arora:213505}] (solid circles), Ref.~[\onlinecite{10.1063/1.2245213}] (solid squares), Ref.~[\onlinecite{petrov_ivanov}] (open circle), Ref.~[\onlinecite{PhysRevB.55.9924}] (open square) and Ref.~[\onlinecite{PhysRevLett.79.3974}] (open triangle). The results of theoretical calculations are shown by lines for the GaAs/Al$_{0.35}$Ga$_{0.65}$As well at $t_{l-h} = 0.5$ (upper panel), and the In$_{0.53}$Ga$_{0.47}$As/InP and CdTe/Cd$_{0.74}$Mg$_{0.26}$Te wells at $t_{l-h}=0$ (lower panel).
The curves are calculated for the free hole (dashed lines) and for the hole bound in an exciton (solid lines). The exciton binding energies used in the calculations are listed in the text.}
\end{figure} 

Let us now apply the theory developed in Sec.~II to describe the experimental data. The experimental data (Tab.~\ref{table1}) and the results of calculations for the InGaAs/InP, CdTe/CdMgTe and GaAs/AlGaAs wells are presented in Fig.~\ref{fig:fig1}. The hole contribution to the exciton $g$-factor is calculated following Eq.~\eqref{eq:exc2} assuming that coefficient $C_x = 1$ and is independent of the well width. Strictly speaking, in Eq.~\eqref{eq:exc2} one should use the binding energy of a two-dimensional exciton $E_{1s}$, however in the calculations we used smaller values of $E_{1s}$ which correspond to realistic wells. Let us note that such a choice of $E_{1s}$ overestimates the second term in Eq.~\eqref{eq:exc2}.

The calculations following Eq.~\eqref{eq:glh1_all} with the use of parametrization (C) and $t_{l-h} = 0$ result for the In$_{0.57}$Ga$_{0.43}$As/InP well in $g_{lh1} > 0$ with an absolute value which is close to the one observed in experiment. Interestingly, $lh1$ level in this well lies at higher energy than $hh2$, therefore the calculations in the multi-level approach yield considerably smaller (and closer to experimental observations) values of $g_{lh1}$ than those calculated in the resonant approximation. Account for excitonic effects in this well ($E_{1s} = 5$~meV~\cite{Mozume2004703}) leads to the decrease of the $g$-factor by approximately 1.5 times. The calculations for the CdTe/Cd$_{0.74}$Mg$_{0.26}$Te well with the use of the set of parameters (D) yield large negative values of $g_{lh1}$. It is known, however, that Coulomb effects in this well are considerable~\cite{KuhnHeinrich:1993uq}, and the calculations with $E_{1s} = 20$~meV give essentially smaller absolute values of the $g$-factor. The top panel of Fig.~\ref{fig:fig1} presents the curves calculated for the GaAs/Al$_{0.35}$Ga$_{0.65}$As well at $t_{l-h} = 0.5$ for the set of parameters (А). The calculations for the free hole are performed following Eq.~(9a) of Ref.~\cite{Durnev2012797} and give overestimated values of the $g$-factor as compared to the experiments. The only way to get satisfactory agreement with the experimental data in this case is to include Coulomb interaction, see the solid black line in Fig.~\ref{fig:fig1} ($E_{1s} = 10$~meV). It is noteworthy that since both calculations for the GaAs well are performed in the resonant approximation it is impossible to extend it in the region of sufficiently thin wells ($a \lesssim 40$~\AA) where the $hh2$ state is already delocalized. The quantitative description of the experimental data in the free hole limit for GaAs/AlGaAs wells requires large values of the interface mixing parameter ($t_{l-h} \approx 3$) which corresponds to the square-root dependence of Zeeman splitting in the whole range of magnetic fields accessed in experiments. Let us note that the parametrization (B) does not allow for qualitative agreement with the experimental data neither for the free hole nor for the exciton. 

\section{Conclusions}
 
To conclude, in this work we have obtained the analytical expressions for the heavy-hole and light-hole $g$-factors in the case of a finite-barrier quantum well taking into account all the size-quantized levels of the hole. It was shown that the interface mixing of the heavy-hole and light-hole states results in the nonlinear behavior of light-hole Zeeman splitting with magnetic field. 
Allowance for Coulomb interaction between an electron and hole recovers linear Zeeman effect. The developed theory with account for excitonic effects allows for satisfactory description of the experimental data on the light-hole $g$-factor in InGaAs/InP, CdTe/CdMgTe and GaAs/AlGaAs wells.

The author is grateful to M. M. Glazov, E. L. Ivchenko and S. A. Tarasenko for valuable discussions. This work was supported by the Dynasty Foundation, RFBR, as well as EU projects SPANGL4Q and POLAPHEN.


\end{document}